\documentclass[sigconf]{acmart}

\AtBeginDocument{%
  \providecommand\BibTeX{{%
    \normalfont B\kern-0.5em{\scshape i\kern-0.25em b}\kern-0.8em\TeX}}}

\copyrightyear{2022} 
\acmYear{2022} 
\setcopyright{acmlicensed}\acmConference[CHIIR '22]{ACM SIGIR Conference on Human Information Interaction and Retrieval}{March 14--18, 2022}{Regensburg, Germany}
\acmBooktitle{ACM SIGIR Conference on Human Information Interaction and Retrieval (CHIIR '22), March 14--18, 2022, Regensburg, Germany}
\acmPrice{15.00}
\acmDOI{10.1145/3498366.3505786}
\acmISBN{978-1-4503-9186-3/22/03}

\begin{document}

\title[Interactive AI for Fact-Checking]{The Effects of Interactive AI Design on User Behavior: An Eye-tracking Study of Fact-checking COVID-19 Claims}

\author{Li Shi}
\orcid{0000-0002-8252-4363}
\affiliation{%
  \department{IX Lab, School of Information}
  \institution{The University of Texas at Austin}
  \city{Austin}
  \state{Texas}
  \country{USA}
}
\email{lilylashi@utexas.edu}

\author{Nilavra Bhattacharya}
\orcid{0000-0001-7864-7726}
\affiliation{%
  \department{IX Lab, School of Information}
  \institution{ The University of Texas at Austin}
  \city{Austin}
  \state{Texas}
  \country{USA}}
\email{nilavra@ieee.org}

\author{Anubrata Das}
\affiliation{%
  \department{School of Information}
  \institution{The University of Texas at Austin}
  \city{Austin}
  \state{Texas}
  \country{USA}}
\email{anubrata@utexas.edu}

\author{Matthew Lease}
\affiliation{%
  \department{School of Information}
  \institution{The University of Texas at Austin}
  \city{Austin}
  \state{Texas}
  \country{USA}}
\email{ml@utexas.edu}

\author{Jacek Gwizdka}
\orcid{0000-0003-2273-3996}
\affiliation{%
  \department{IX Lab, School of Information}
  \institution{The University of Texas at Austin}
  \city{Austin}
  \state{Texas}
  \country{USA}}
\email{jacekg@utexas.edu}

\renewcommand{\shortauthors}{Shi, et al.}

\begin{abstract}
  We conducted a lab-based eye-tracking study to investigate how interactivity of an AI-powered fact-checking system affects user interactions, such as dwell time, attention, and mental resources involved in using the system. 
  A within-subject experiment was conducted, where participants used an interactive and a non-interactive version of a mock AI fact-checking system, and rated their perceived correctness of COVID-19 related claims. 
  We collected web-page interactions, eye-tracking data, and mental workload using NASA-TLX. We found that the presence of the affordance of interactively manipulating the AI system's prediction parameters affected users' dwell times, and eye-fixations on AOIs, but not mental workload. 
  In the interactive system, participants spent the most time evaluating claims' correctness, followed by reading news. 
  This promising result shows a positive role of interactivity in a mixed-initiative AI-powered system. 
\end{abstract}



\maketitle

\section{Introduction}
As an important task in Information Retrieval (IR), fact-checking has various implications for both system-centered and user-centered IR.
These include which results should be returned, how they should be presented, what models of interaction should be provided, and how can success be evaluated. 
Many new models for automatic fact-checking of claims were recently developed in machine learning and natural language processing literature \cite{hassan2015quest,hassan2017toward,jiang2020factoring,zubiaga2018detection}.
These models, however, focused on fully-automated fact-checking, and maximizing model’s predictive accuracy.

While accurate predictions are important, a user doubtful of online information is likely to remain skeptical of any fact-checking tool \cite{kutlu2018mix}.
As with all AI systems, fact-checkers operate on limited resources and thus are failable and prone to errors. Users may arrive at a wrong decision influenced by model errors.
Nguyen et al. \cite{nguyen2018believe} show that users might trust a fact-checking model even when the model is wrong.
Further, Mohseni et al. \cite{mohseni2020machine} show that transparent systems prevent users from over-trusting model predictions. 
Effective human-AI teaming \cite{bansal2019beyond} might alleviate such issues in fact-checking models for which a system needs to be transparent. 
A transparent system could reveal to a user how it made a prediction to support a user’s understanding and calibrate trust \cite{amershi2019guidelines,bansal2019beyond,ferreira2020people}. 
Moreover, individual claim assessments will certainly partially rely on the user’s prior worldviews concerning the perceived credibility of sources and claims.

Thus, a fact-checking system needs to integrate user beliefs, and enable users to infuse their views and knowledge into the system.
Additionally, such a system needs to be transparent, by communicating the prediction-uncertainty of its model and enabling users to perform their in-depth reasoning.  
Recent investigations have also suggested that search engines, like any technology, have the potential to harm their users \cite{azzopardi2021cognitive,kattenbeck2019understanding,pogacar2017positive}.
Allowing users to interact with a tool might help in mitigating such harm. 

In this short paper, we present an eye-tracking and usability analysis of a human-AI mixed-initiative fact-checking system. 
We believe that such systems can potentially augment the efficiency and scalability of automated IR, with transparent and explainable AI \cite{kutlu2018mix,nguyen2018believe,nguyen2018interpretable}. 
Starting with a claim as a query, the system retrieves relevant news articles. 
It then infers the degree to which each article supports or refutes the claim (stance), as well as each news source’s reputation. 
The system then aggregates this evidence and predicts the claim correctness. 
It shows to the user how the information is being used, and what are the sources of the model’s uncertainty. 

\begin{figure*}[h]
  \centering
  \includegraphics[width=\linewidth]{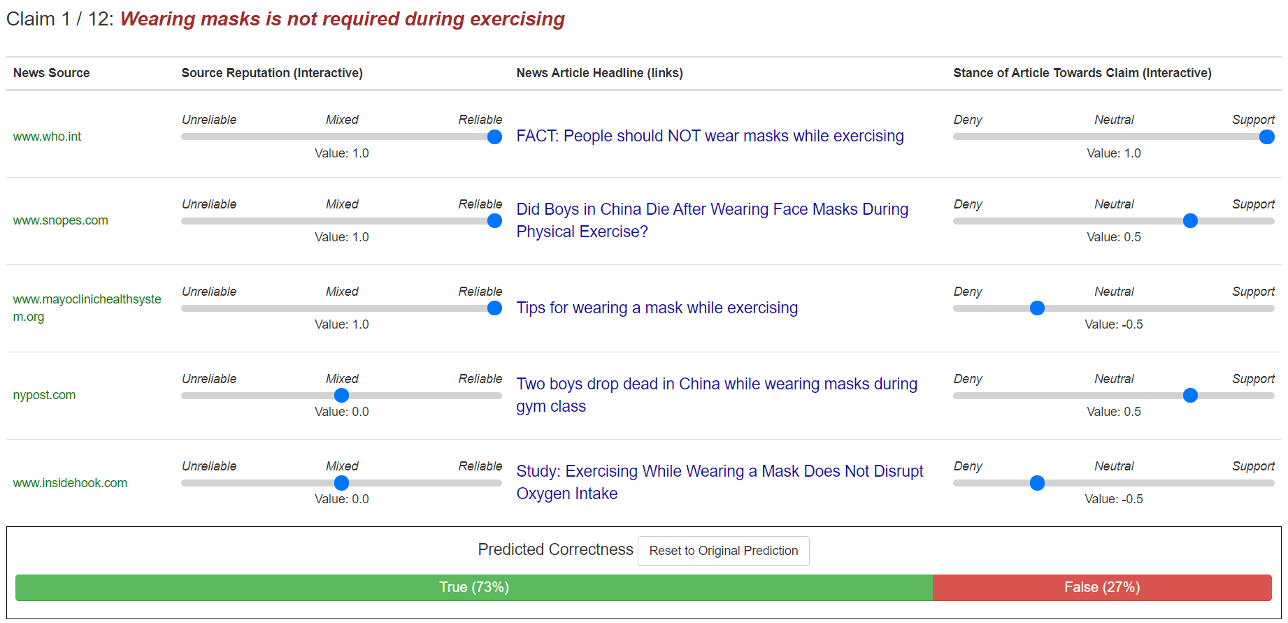}
  \caption{Interactive version of the fact-checking system showing different areas of interest (AOIs)}
  \Description{A fact-checking system UI with manipulation function and demonstrates different AOIs on the screen}
  \label{fig:interactive_ui}
\end{figure*}

\begin{figure}[h]
  \centering
  \includegraphics[width=\linewidth]{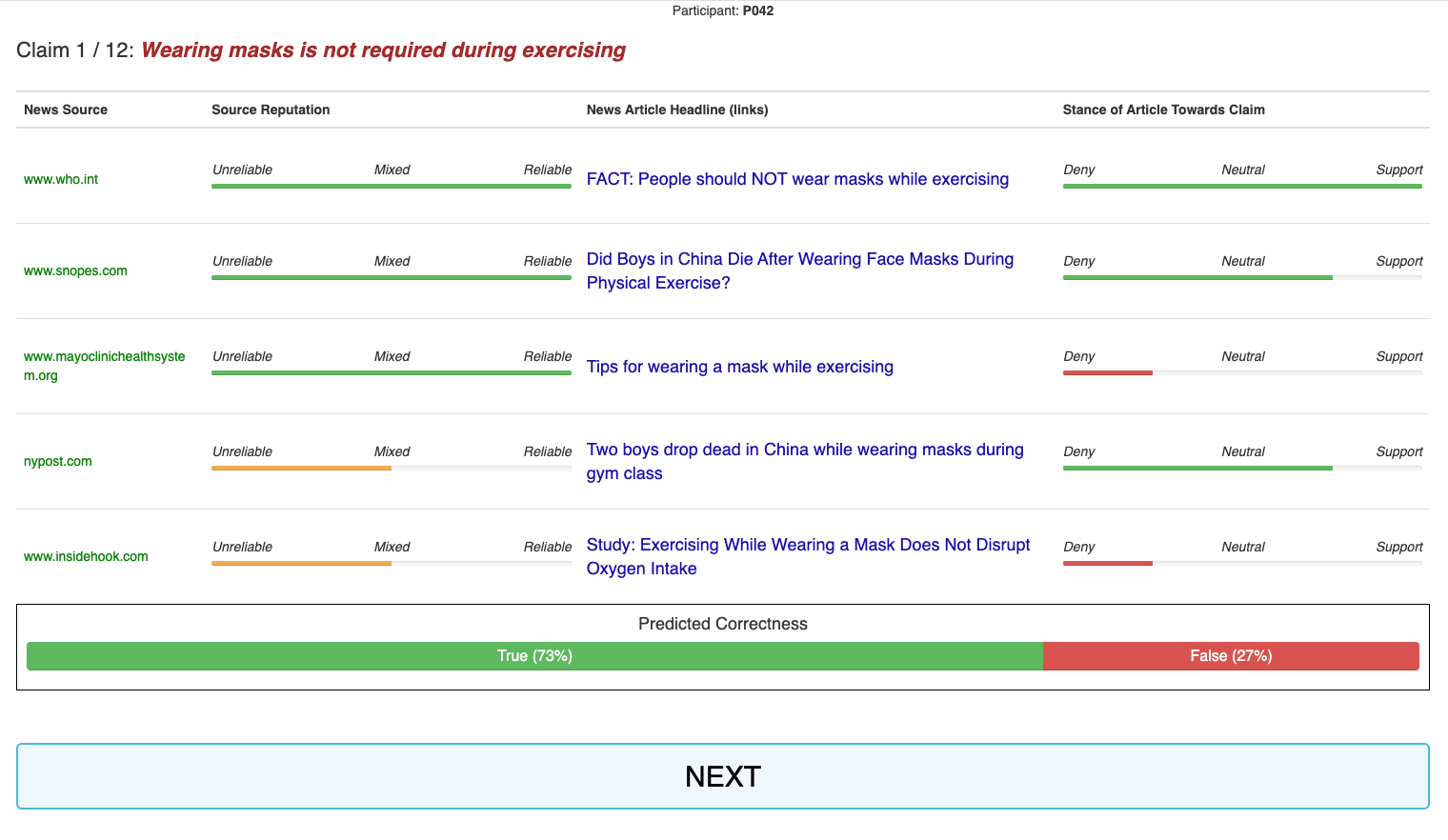}
  \caption{Non-interactive version of the system}
  \Description{A fact-checking system UI without manipulation function.}
  \label{fig:non_interactive_ui}
\end{figure}

Our focus in this paper is not to evaluate the fact-checking IR model per se, but rather to understand how users interact with the system via two different user interfaces: an interactive interface and a non-interactive interface. 
In the interactive interface (Figure ~\ref{fig:interactive_ui}), the model's predicted source reputation and stance for each retrieved article, is shown to the user. These can be revised via simple sliders to reflect a user’s beliefs and/or to correct erroneousness model estimates. 
The overall claim prediction is then updated visually in real-time, as the user interacts with the system. In the non-interactive interface (Figure ~\ref{fig:non_interactive_ui}), the values of source bars, with no option to change any of the data presented.

Therefore, the key research aim is to study the effects of the interactive AI design on user behavior in fact-checking systems. 
This paper investigates three aspects of the user behavior: dwell time, attention, and mental workload. 
We hypothesize that people would spend more time in the interactive interface, pay more attention to the interactive elements, and that the interactive interface would impose higher mental workload.

\section{Method}
\subsection{Experimental design}
A controlled, within-subjects eye-tracking study was conducted in the Information eXperience usability lab at the University of Texas at Austin (N=40, 22 females).
Voluntary participants interacted with two versions (user-interfaces) of a fact-checking system. 
Participants were pre-screened for native-level English familiarity, 20/20 vision (uncorrected or corrected), and non-expert topic familiarity of the content being shown.
A Tobii TX-300 eye-tracker was used to record the participants’ eye movements. Upon completion of the experiment, each participant was compensated with USD 25.

\subsection{Tasks}

Each participant interacted with two versions (user-interfaces) of a fact checking system. 
In each interface, there were 12 trials. 
Each trial consisted of viewing a claim and, optionally, its corresponding news articles. 
Screenshots of single trials from the two versions of the system are shown in Figures 1 and 2. 
In each trial, a claim was shown at the top of the interface, and surrogates of five related news articles were presented below, each with its corresponding news source, source reputation, news headline, and the article’s stance towards the claim. 
Based on the article’s stance and news source reputation, the system provided a prediction of the claim’s correctness at the bottom. 
The news headlines were clickable and upon clicking opened the news article in a new tab.

The claims and corresponding news-articles were on the topic of the COVID-19 pandemic. 
They were handpicked by the researchers to simulate a mock version of the fact-checking system for usability analysis. 
Each claim was selected so as to have a pre-assigned ground-truth correctness value of TRUE, FALSE, or UNSURE (claims that are partially true, or not totally proven).
The TRUE and UNSURE claims were handpicked from reputed websites in the medical domain, such as World Health Organization, WebMD, Mayoclinic, Johns Hopkins University, US State Government webpages, and others. 
The FALSE claims were selected by searching for “coronavirus myths” on popular search engines. 
The supporting news articles for each claim were collected by manually searching the web. 
The source reputations for news articles were collected from existing datasets \cite{gruppi_nela-gt-2019_2020,norregaard2019nela}, while the stance values of each news article towards each claim was labelled by the researchers. 
Two example claims are “wearing masks is not required during exercising”, and “asymptomatic people can transmit COVID-19”. 
In total there were 24 claims (8 TRUE, 8 FALSE, 8 UNSURE), distributed equally between both interfaces. 
The order of the interfaces (interactive / non-interactive) was balanced, and the order of the claims were randomized. 
The list of claims and corresponding news articles used are shared in the GitHub repository: \url{https://github.com/ixlab-ut/chiir-2022}.

\subsection{Procedure}

\begin{figure}[h]
  \centering
  \includegraphics[width=\linewidth]{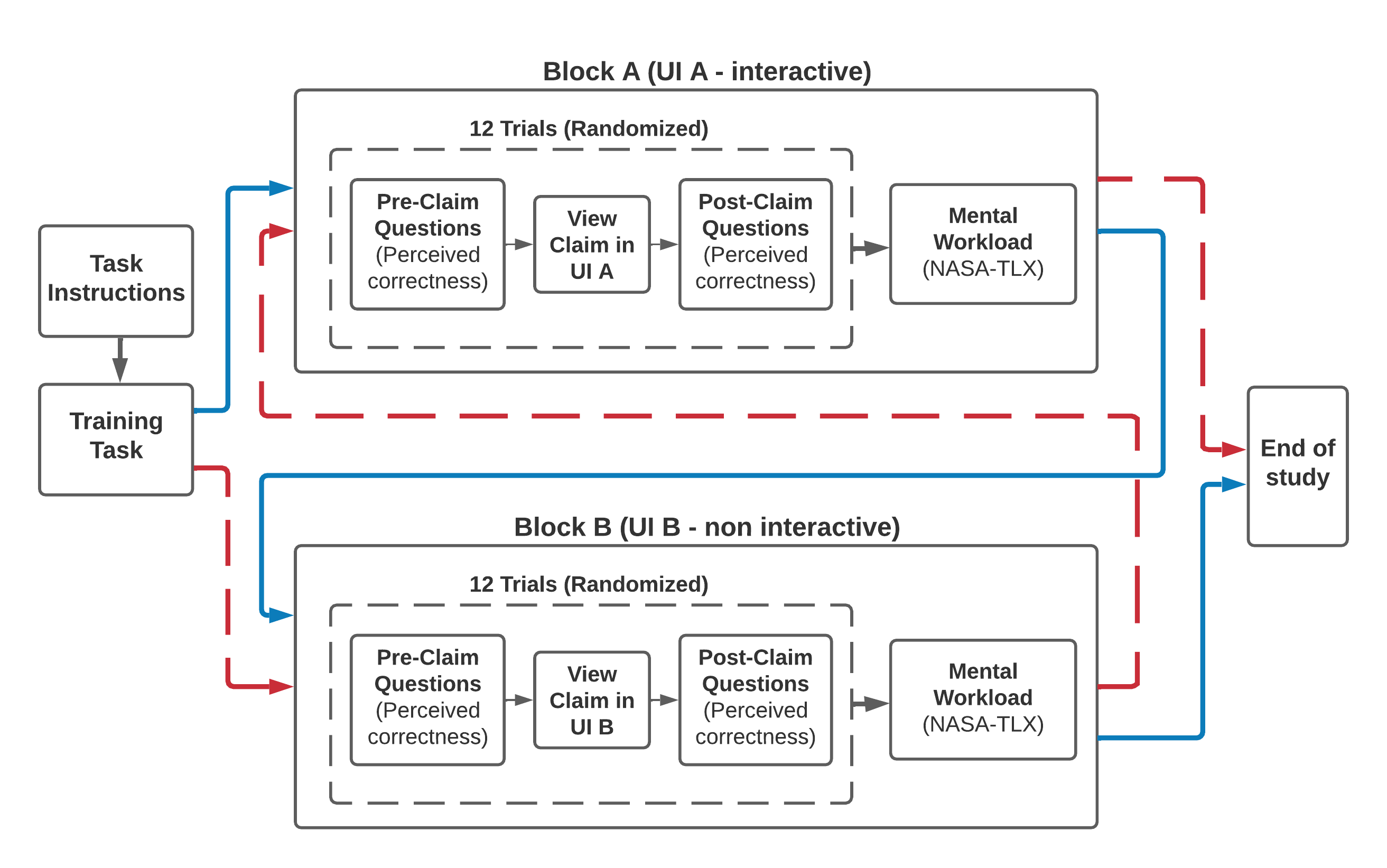}
  \caption{Experimental Procedure}
  \Description{Flowchart of the experiment.}
  \label{fig:experiment_flowchart}
\end{figure}

The overall procedure of the experimental session is illustrated in Figure ~\ref{fig:experiment_flowchart}. 
Each session started with two training trials (one in interactive, one in non-interactive) for participants to get familiar with the two fact-checking interfaces. 
Then the participants started trials in one of the two interfaces (experiment blocks), which was randomly chosen and balanced across all participants. 
In each trial (viewing a claim) participants interacted with the interface freely without a time limit.
Participants were also instructed to click on news headlines to open the underlying news articles in a new browser tab, and read it, if they considered it necessary for evaluating the claim. 
Before and after viewing each claim in the system, participants indicated their perceived-correctness of the claim (which is not analyzed in this short paper). 
After completing 12 trials in the first interface (block), participants reported their mental workload using the NASA-TLX questionnaire \cite{hart2006nasa}. 
Then they were allowed to take a five-minute break before resuming the second block (in the other interface).
NASA-TLX questionnaire was again administered at the end of the block.

\subsection{Measures}
\subsubsection{Eye-tracking (ET) Measures}
To study the effect of the differences in design of the user-interfaces on user’s information behavior, we collected eye-tracking data to reflect the participants’ attention, reading, and information processing.
We divided the interfaces into six areas of interests (AOIs): (i)  claim text (T); (ii) news source names (S); (iii) source reputation (R); (iv) news article headlines (H); (v) stance of news article towards the claim (A); and (vi) predicted correctness of the claim (C). 
The ET measures comprise total fixation count, total fixation duration, and average fixation duration. 
For analysis, we discarded low-quality and non-reading eye tracking data, which are the fixations with durations shorter than 100 milliseconds or longer than 1500 milliseconds \cite{holmqvist2011eye}.

\subsubsection{Interface and News Dwell Times}
To assess how much time participants spent in interacting with our system vis a vis how much time they spent on reading the original news articles (which opened in a new browser tab upon clicking the news headlines in the interface), we recorded the total dwell times while participants were viewing the interface (interface dwell time), and when they were reading news articles (news dwell time).

\subsubsection{Mental Workload}
We used NASA-TLX questionnaire to measure mental workload after task completion in each interface use. The questionnaire includes six questions which measure six workload factors.

\section{results}
\subsection{Dwell time}
\begin{figure}[h]
  \centering
  \includegraphics[width=\linewidth]{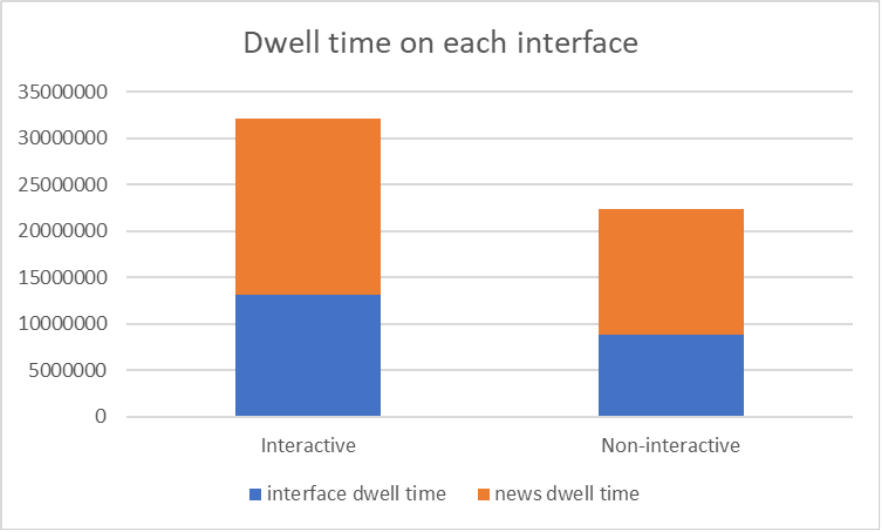}
  \caption{Dwell time on the two interfaces (milliseconds).}
  \Description{Dwell time chart. Interface dwell time is smaller than news dwell time portion.}
  \label{fig:dwell_time}
\end{figure}

As shown in Figure ~\ref{fig:dwell_time}, the participants generally spent more time reading news than using the interface. 
Overall, in the interactive interface, participants spent more time using the fact-checking system (\textit{interface dwell time}), as well as reading the underlying news articles (\textit{news dwell time}).

A Wilcoxon signed ranks test (T=255, n=40, p<.05) indicated that the \textit{total dwell time} spent on reading the underlying news articles was significantly different between the two interfaces. 
The sum of the positive difference ranks (\(\sum R_{+}= 564\)) was larger than the sum of the negative difference ranks (\(\sum R_{-}= 255\)), showing that people spent more time reading news in the interactive interface, than in the non-interactive interface. 
The effect size for the test was 0.33, signifying that presence of interactivity in the system had an effect on making users spend more time reading underlying news articles.

\subsection{Fixation count and duration}

\begin{figure}[h]
  \centering
  \includegraphics[width=\linewidth]{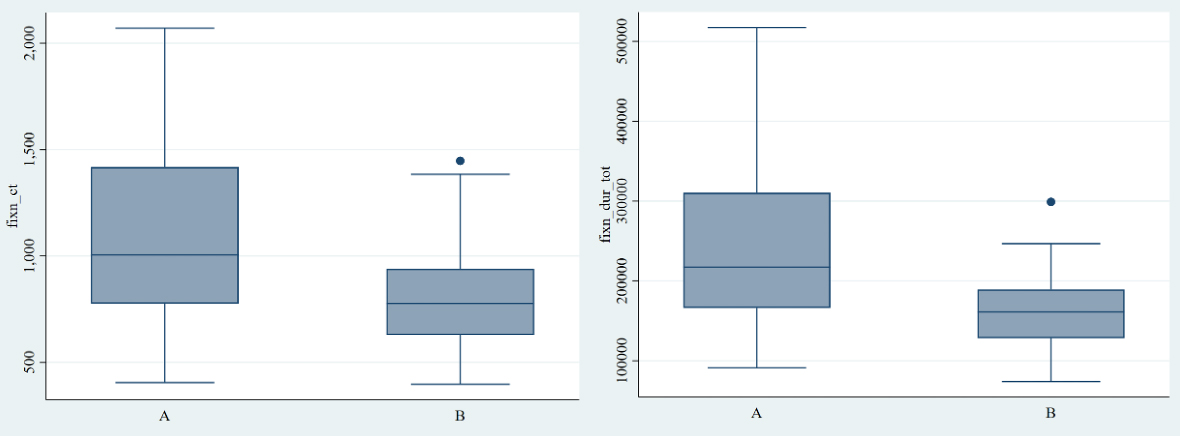}
  \caption{Total fixation count (left), total fixation duration (right) in interactive (A) and non-interactive (B) interfaces.}
  \Description{Box chart of fixation counts and fixation duration in different interfaces.}
  \label{fig:fixation_ui}
\end{figure}

\textbf{UI and ET.}
Figure ~\ref{fig:fixation_ui} shows that \textit{total fixation count} and \textit{total fixation duration} have wider spread and higher maximums in the interactive interface. 
We employed the Wilcoxon signed ranks between the two interfaces to test for significant differences in \textit{total fixation count} (T=99, n=40, p<.05) and \textit{total fixation duration} (T=77, n=40, p<.05). 
The results indicated that the fixation measures differed significantly between the two interfaces. 
In addition, the sums of the positive difference ranks for the \textit{total fixation count} (\(\sum R_{+}=721\)) and \textit{total fixation duration} (\(\sum R_{+}=743\)) were larger than the sums of the negative difference ranks for the \textit{total fixation count} (\(\sum R_{-}=99\)) and \textit{total fixation duration} (\(\sum R_{-}=77\)), respectively. 
Therefore, there were more fixations and overall longer eye-dwell time in the interactive interface. Moreover, the effect size for the matched-pair samples was 0.66 for \textit{total fixation count} and 0.71 for \textit{total fixation duration}, showing that presence of interactivity had a strong effect of attracting participants’ visual attention and processing. 
We did not find any significant differences in \textit{average fixation duration}.

\begin{figure}[h]
  \centering
  \includegraphics[width=\linewidth]{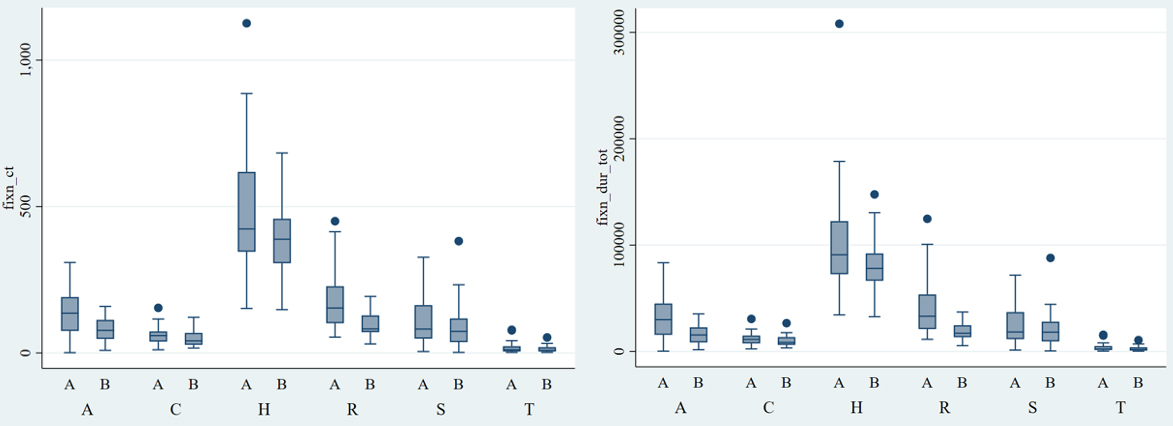}
  \caption{Total fixation count (left), total fixation duration (right) for different AOIs in two interfaces. (A: stance of article, C: predicted correctness, H: news article headlines, R: source reputation, S: news source, T: claim text).}
  \Description{Boxplot of fixation counts and duration in different AOIs and interfaces.}
  \label{fig:fixation_ui_aoi}
\end{figure}

\textbf{AOIs, UI and ET.} 
Figure ~\ref{fig:fixation_ui_aoi} illustrates that most fixations and longer fixation durations were on the ‘news article headlines’ area of interest (AOI), followed by the ‘source reputation’, ‘news source’, and ‘stance of article’ AOIs.
The ‘predicted correctness’ AOI had fewer fixations and shorter durations, while the ‘claim text’ AOI had the least.

A two-way ANOVA was conducted to examine the effect of interface and AOI on eye-fixation metrics. There was a significant interaction between the effects of interface and AOI on \textit{total fixation count}, F(5, 464) = 4.00, p < .05, \textit{total fixation duration}, F(5, 464) = 4.41, p < .05, and \textit{average fixation duration}, F(5, 464) = 3.42, p < .05. Simple main effects analysis showed that both interface and AOI had significant main effects on all three of the eye-fixation metrics.

\begin{figure}[h]
  \centering
  \includegraphics[width=\linewidth]{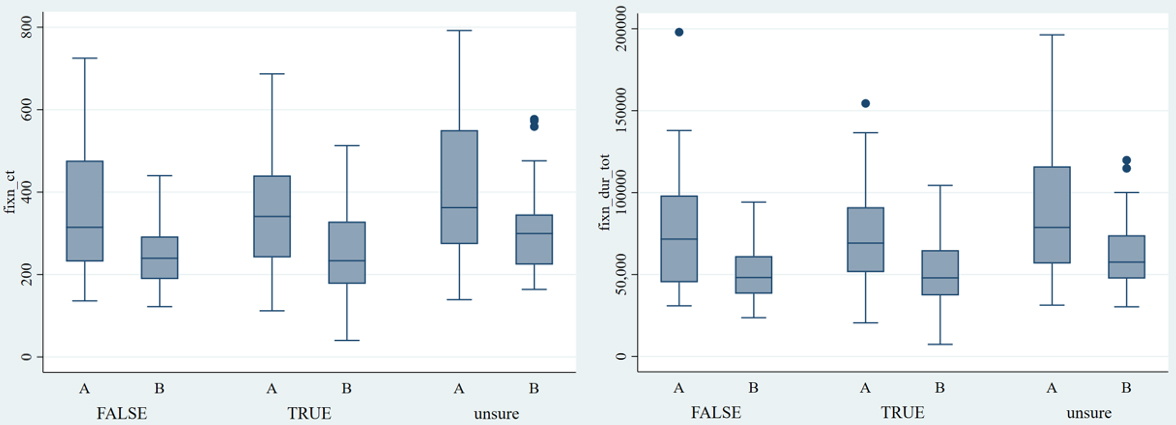}
  \caption{Total fixation count (left) and total fixation duration (right) for different claim conditions (TRUE / FALSE / UNSURE) in different interfaces (A=interactive, B = non-interactive).}
  \Description{Boxplot of fixation duration and fixation counts of different claim correctness condition and interfaces.}
  \label{fig:fixation_ui_et}
\end{figure}

\textbf{Claim correctness, UI and ET.} 
Figure ~\ref{fig:fixation_ui_et} shows the eye-tracking metrics for viewing TRUE, FALSE, and UNSURE claims in the two interfaces. 
The \textit{total fixation count} and \textit{total fixation duration} while viewing UNSURE claims are higher than when viewing TRUE or FALSE claims. 
To investigate further, a two-way ANOVA was conducted to examine the effect of interface and claim-correctness on eye-fixation metrics. Interface type had a significant main effect on all three eye-fixation metrics: \textit{total fixation count}, F(1, 234) = 37.42, p < .05, \textit{total fixation duration}, F(1, 234) = 42.18, p < .05, and \textit{average fixation count}, F(1, 234) = 4.85, p < .05. 
Claim correctness had a significant main effect on \textit{total fixation count}, F(2, 234) = 5.95, p < .05, and \textit{total fixation duration}, F(2, 234) = 5.50, p < .05, but not on \textit{average fixation duration}. 
Interaction effects were also not significant.

\subsection{Mental workload}
The Wilcoxon signed ranks test (T=208, n=40, p>.05) indicated that the participants’ mental workload was not significantly different between the two interfaces. Therefore, the interactivity did not significantly influence the mental workload level.

\section{Discussion and Conclusion}
We conducted a lab-based experiment to investigate how interactivity of an AI-powered fact-checking system affects user interactions. 
We found that the interactivity of the system has an influence on the system dwell time, fixation count and duration, but not mental workload.

Overall people tended to spend more time on reading the original news than looking at and interacting with the two systems. 
This indicated that they did not rely on the system unilaterally, but read the original news to help them make informed decisions. 
Furthermore, people engaged more and spent more time reading the news in the interactive than in the non-interactive system. 
We found that the interactivity of the interface makes a difference in the fixation counts and durations regardless of the AOI type and claim condition. 
People always paid more attention to the interactive interface when using the fact-checking system. The news headlines drew the most attention among all the interface elements. 
By reading the headlines, people decided which news to view.  
We also found that the fixation counts and fixation duration differed significantly between the news headlines, news source, source reputation, and article stance. 
Users tended to pay more attention to them, when the interface elements were interactive. 
The difference between the claim conditions showed that people paid more attention to the UNSURE claims. 
They needed more information from the system when they were dealing with the UNSURE claims. People’s mental workload was not influenced by the system interactivity.
Even though they paid more attention to the interactive interface, the amount of perceived mental resources required was apparently not significantly changed. 
Thus, the interactivity did not increase the self-reported effort of using the system.

These findings reveal that the system interactivity plays a positive role in a mixed-initiative AI-powered system.
The system interactivity encourages people to spend the most time evaluating the claim correctness and then reading the news, while not imposing higher mental workload on users.

Limitations of our work include using only researcher-assigned tasks and claims selected by researchers, which were assessed by participants in the lab environment rather than in the context of their natural information tasks and fact-checking.
Additionally, we did not capture participants’ prior knowledge regarding the particular claims in the study.
In practice, fact-checking is often performed with constrained time and high error costs. 
However, this study does not incentivize the participants for accuracy and efficiency.

Future work will include using sets of claims on different topics and investigation of user interaction in the context of their naturally generated tasks.
Surprisingly, we notice that users spend a significant chunk of time reading the linked articles. 
Future work could explore the effect of time constraints and incentivizing participants’ accuracy on engagement with interactive AI. 
Specifically, a time-constrained setup might encourage participants to engage with the AI outcomes more instead of reading all the articles themselves. 
We see that participants pay more attention to the interactive interface. 
However, it is unclear whether this increased attention stems from the apparent correlation between interactivity and the attention required for engagement with interactive interfaces. 
Future work could bolster the current findings by closely observing prolonged user interaction with the tool.

\begin{acks}
We thank the reviewers for their valuable feedback, and the research participants for their time. This research was completed under UT Austin IRB study 2017070049 and supported in part by Wipro, the Micron Foundation, the Knight Foundation, and by Good Systems\footnote{http://goodsystems.utexas.edu/} , a UT Austin Grand Challenge to develop responsible AI technologies. The statements made herein are solely the opinions of the authors and do not reflect the views of the sponsoring agencies.
\end{acks}

\bibliographystyle{ACM-Reference-Format}
\bibliography{references}

\end{document}